\documentclass[10pt,twocolumn,twoside,english]{IEEEtran}

\usepackage{url}
\usepackage{graphicx,epsfig}
\usepackage{caption}
\usepackage{subcaption}
\usepackage{cite}
\usepackage{stfloats}
\usepackage{amsmath}
\usepackage{amsfonts,helvet}

\usepackage{algorithm}
\usepackage{algorithmic}
\usepackage{multirow}
\usepackage{babel}
\makeatletter
\adddialect\l@ENGLISH\l@english
\makeatother

\makeatletter

\let\proof\@undefined
\let\endproof\@undefined
\makeatother

\usepackage{amsthm}

\newtheorem{theorem}{Theorem}

\newtheorem{algo}[theorem]{Algorithm}

\newtheorem{proposition}[theorem]{Proposition}

\newcounter{proposition}
\begin{document}

\title{Compressed Channel Feedback for Correlated Massive MIMO Systems}

\author{Min~Soo~Sim,~\IEEEmembership{Student~Member,~IEEE}, Jeonghun~Park,~\IEEEmembership{Student~Member,~IEEE}, Chan-Byoung~Chae,~\IEEEmembership{Senior~Member,~IEEE}, and Robert~W.~Heath,~Jr.,~\IEEEmembership{Fellow,~IEEE} \\
\thanks{M. S. Sim and C.-B. Chae are with the School of Integrated Technology, Yonsei University, Korea (E-mail: \{simms, cbchae\}@yonsei.ac.kr). J. Park and R. W. Heath, Jr. are with the Dep. of ECE, Univ. of Texas at Austin, USA (E-mail: \{Jeonghun, rheath\}@utexas.edu). This work was supported by the the Ministry of Knowledge Economy under the IT Consilience Creative Program (NIPA- 2014-H0201-14-1002). Part of this work was presented in~\cite{sim2014compressed}.
 }}




\maketitle \setcounter{page}{1} 
%
%
%

\markboth{Sim \emph{{et al.}}: Compressed Channel Feedback for Correlated Massive MIMO Systems}
{Sim \emph{{et al.}}: Compressed Channel Feedback for Correlated Massive MIMO Systems}

\begin{abstract}
Massive multiple-input multiple-output (MIMO) is a promising approach for cellular communication due to its energy efficiency and high achievable data rate. These advantages, however, can be realized only when channel state information (CSI) is available at the transmitter. Since there are many antennas, CSI is too large to feed back without compression. To compress CSI, prior work has applied compressive sensing (CS) techniques and the fact that CSI can be sparsified. The adopted sparsifying bases fail, however, to reflect the spatial correlation and channel conditions or to be feasible in practice. In this paper, we propose a new sparsifying basis that reflects the long-term characteristics of the channel, and needs no change as long as the spatial correlation model does not change. We propose a new reconstruction algorithm for CS, and also suggest dimensionality reduction as a compression method. To feed back compressed CSI in practice, we propose a new codebook for the compressed channel quantization assuming no other-cell interference. Numerical results confirm that the proposed channel feedback mechanisms show better performance in point-to-point (single-user) and point-to-multi-point (multi-user) scenarios.
\end{abstract}



\section{Introduction}
\label{intro}

The concept of multiple-input multiple-output (MIMO) wireless communication employing a number of antennas, a.k.a. massive MIMO, has been researched for several years. It was found that a base station (BS) with more antennas can recover information in lower signal-to-noise-ratio (SNR) when the number of antennas is sufficiently large~\cite{marzetta2006how}. With this motivation, the idea of using a very large number of antennas at the BS in a cellular system was proposed in~\cite{marzetta2010noncooperative}. Massive MIMO systems are known to provide large network capacity gain by supporting many users~\cite{hoydis2013massive}, and higher energy efficiency~\cite{ngo2011energy}. Practical issues, transmit precoding and receive post processing, and channel estimation issues for massive MIMO systems were discussed in~\cite{rusek2013scaling, Lim2015performance}.

A transmitter with multiple antennas has to exploit channel state information (CSI) to provide beamforming gains in single-user (SU) MIMO systems, and multiplexing gains in multi-user (MU) MIMO systems~\cite{gesbert2007shifting}. With inaccurate CSI, however, there is sum-rate saturation even in massive MIMO systems~\cite{jindal2006mimo, ding2007multiple}. It is, therefore, important to design efficient channel estimation and feedback strategies. In time division duplexing (TDD) systems, CSI can be implicitly obtained using reciprocity. In frequency division duplexing (FDD), which most of cellular systems employ nowadays, the receiver has to feed back information of channel state or precoding vectors. It is known that the feedback overhead must increase to maintain a certain level of CSI quantization loss~\cite{hassibi2003how, yeung2007onthe, chae2008CBFJSAC, chae2008CBFTSP}. From this point-of-view, it is essential to compress and quantize CSI efficiently due to the large number of antennas. To solve these issues, a feedback reduction technique that exploits spatial correlation of users was proposed in~\cite{nam2012joint}, and noncoherent trellis-coded quantization for FDD massive MIMO systems was proposed in~\cite{choi2013noncoherent}. In~\cite{nam2012joint, choi2013noncoherent}, however, it was assumed that the spatial correlation matrices are perfectly available at transmitters.

Compressive sensing (CS) based CSI compression was applied in~\cite{kuo2012compressive}. It uses the fact that CSI in massive MIMO systems has high spatial correlation due to the limited physical distance between antennas. The theory of CS~\cite{Candes:rip, donoho2006compressed, baraniuk2007compressive} has been applied in various areas including signal processing and communications, where the information is sparse. A sparse signal (or vector) is a signal that can be represented by few elements in a certain domain. Via random projections, CS is able to compress sparse information efficiently. With the insight that CSI can be represented in sparse form in a spatial-frequency domain, two sparsifying bases were adopted in~\cite{kuo2012compressive}: the two-dimensional discrete cosine transform (2D-DCT) and the instantaneous Karhunen-Loeve transform (KLT). Unlike~~\cite{nam2012joint, choi2013noncoherent}, there is no need to assume transmitters to know the correlation matrices in~\cite{kuo2012compressive}. Without this assumption, however, the 2D-DCT basis fails to reflect the spatial correlation of the systems. The instantaneous KLT basis changes as the channel varies, making it, in practice, unfeasible. CS techniques simplify encoding, but require solving an optimization problem for decoding, thus demanding significant computing resources.

In this paper, we propose two new compression methods for channel feedback in massive MIMO systems using the fact that highly correlated CSI can be represented in a sparse form. For a sparsifying basis, we adopt the KLT, which considers the long-term correlation model of the channel. The first method compresses via random projection, while the second one uses the sparsifying basis directly. The former method is useful when the receiver does not know what basis to use (Scenario~1), while the latter method is prefered when the receiver and the transmitter select what sparsifying basis to use (Scenario~2). To quantize the compressed CSI, 
we adopt the widely used Linde, Buzo, and Gray (LBG) algorithm~\cite{linde1980algorithm}, and random vector quantization (RVQ)~\cite{AuYeung2007perfomanceRVQ, Santipach2003Asymptotic}. The main contributions of this paper are as follows:

\begin{itemize}

\item \emph{Compression method for Scenario~1}:
Since the calculation of the KLT basis from the covariance matrix entails high complexity, the receiver might be unable to obtain the basis. In this case, CS technology that needs no information of the basis for compression but just compresses via random projections is applied. With the 2D-DCT or KLT basis, the indices of the dominant elements of the sparsified CSI are expected to be in certain region. Using this fact, our contribution in random projection-based compression is a new reconstruction algorithm with less complexity compared to conventional decoding algorithms. We show numerically that, compared to the conventional reconstruction method for CS-based compression, channel feedback with the proposed decoding algorithm performs better in terms of recovery accuracy, and achievable rate.

\item \emph{Compression method for Scenario~2}:
In Scenario~2, the transmitter and the receiver can choose what sparsifying basis they will use. If both the transmitter and the receiver have enough computing resourses to obtain the KLT basis, the KLT basis is adopted for sparsifying, and if not, the 2D-DCT basis is adopted. Thus, both the transmitter and the receiver can know the basis without any additive coordination. With the sparsified CSI, we propose to compress it by dimensionality reduction. The proposed method is simpler to compress and reconstruct when the position of the dominant elements in the sparsified CSI is expected to be focused on certain region. We show numerically that, compared to the CS-based compression, the proposed dimensionality-reduction-based compression performs better regarding recovery accuracy, and achievable rate.

\item \emph{Codebook construction}: 
For linear precoding, the Grassmannian codebook has been widely used~\cite{love2003grassmannian}, mostly, in single-user MIMO scenarios. The Grassmannian codebook, however, can only cover one-norm vectors, failing to fit the compressed CSI from either compression method for MU MIMO scenarios. Therefore, we adopt the LBG algorithm, which exploits the statistical properties of the compressed CSI, to generate a codebook. We analyze how the compressed CSI vectors are distributed and construct a codebook based on our analysis.

\end{itemize}

To simplify analysis, we assume that the receiver can estimate perfect CSI without any noise and/or other-cell interference, and that there is an ideal control channel that can send, without errors, the compressed CSI. Also, we assume that spatial correlation is obtainable at the transmitter or at the receiver according to each scenario with no error. This paper is organized as follows. In Section~\ref{model}~and~\ref{background}, we introduce the system model for massive multi-user MIMO systems, and a review of sparse signal compression including CS and dimensionaltiy reduction. In Section~\ref{feedback}, we explain the sparsifying bases, and the details of the compression methods with given bases. We also introduce a codebook generation rule. Performance analysis and our conclusion are given in Sections~\ref{analysis}~and~\ref{conclusion}. 

\begin{figure}[!t]
        \centering
        \begin{subfigure}[b]{0.49\columnwidth}
                \centering
                \includegraphics[width=1.0\columnwidth]{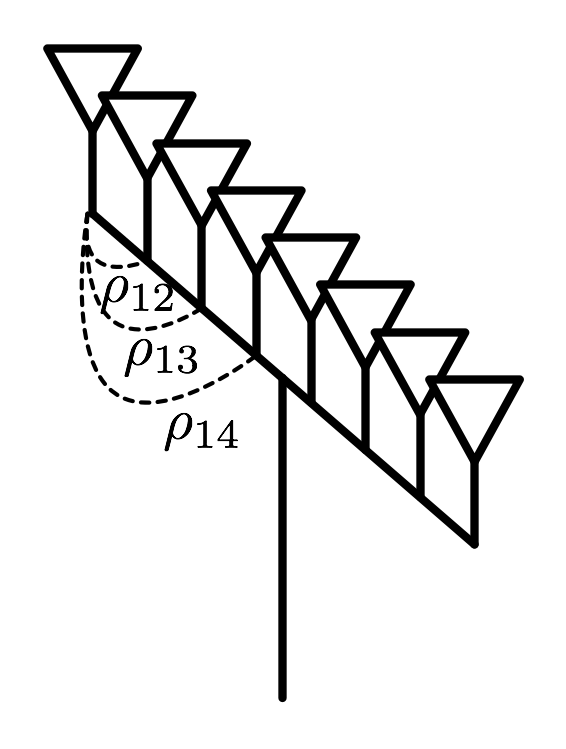}
                \caption{One-dimensional ULA}
                \label{Fig:linear_array}
        \end{subfigure}                
        \begin{subfigure}[b]{0.49\columnwidth}
                \centering
                \includegraphics[width=1.0\columnwidth]{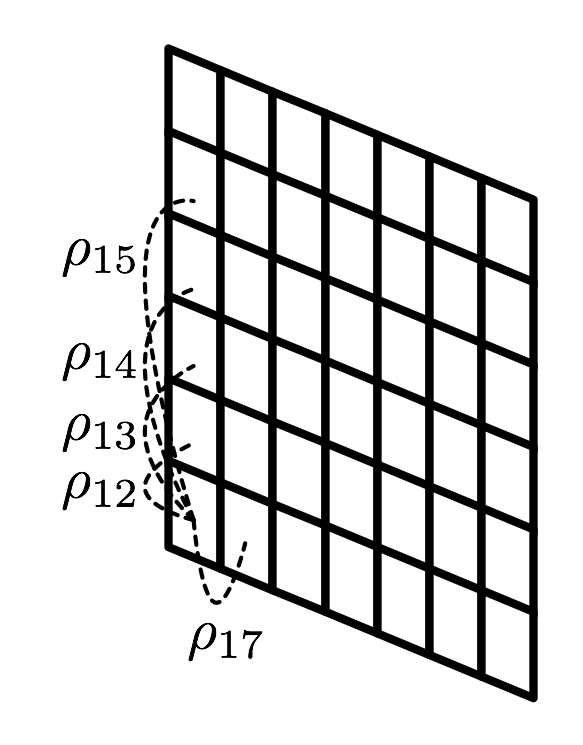}
                \caption{Two-dimensional UPA}
                \label{Fig:square_array}
        \end{subfigure}
        \caption{Geometry and correlations of an ULA and an UPA.}
        \label{Fig:antenna_array}
\end{figure}

\section {System Model}

\label{model}

In this section, we explain the system model and the assumptions.\footnote{Throughout this paper, we use upper and lower case boldface to describe matrix $\pmb{A}$ and vector $\pmb{a}$, respectively. The transpose and the Hermitian transpose of a matrix is notated as $(\cdot)^T$ and $(\cdot)^*$, respectively. The $\text{vec}(\cdot)$ operator stacks the columns of a matrix into a vector. $\mathbb{E}[\cdot]$ denotes the expectation operator. $\otimes$ denotes Kronecker product.}
Consider a MIMO broadcast signal model with $N_\mathrm{u}$ receivers with $N_\mathrm{r}$ receive antennas. Each user receives its own data stream, which is precoded at the transmitter with $N_\mathrm{t}$ antennas. We consider two types of antenna arrays: an one-dimensional uniform linear array (ULA) model, and a two-dimensional uniform planar array (UPA) model. Figure~\ref{Fig:antenna_array} illustrates how arrays are designed. In Figure~\ref{Fig:antenna_array}, we note that the correlation decreases with the distance. Note that our algorithms work well regardless of channel correlation models. In this paper, we do not consider Doppler.

For the one-dimensional ULA model, the 3GPP Spatial Channel Model (SCM) is adopted \cite{3gpp2003scm}. To obtain the KLT basis $\pmb{\Psi}_{\textrm{KLT}}$, $1000$ channel vectors $\pmb{h}$ are generated to calculate the covariance $\pmb{C}_{\pmb{h}}$ for an each transmitter-receiver-link.

\begin{figure*}[!t]
  \centerline{\resizebox{2.0\columnwidth}{!}{\includegraphics{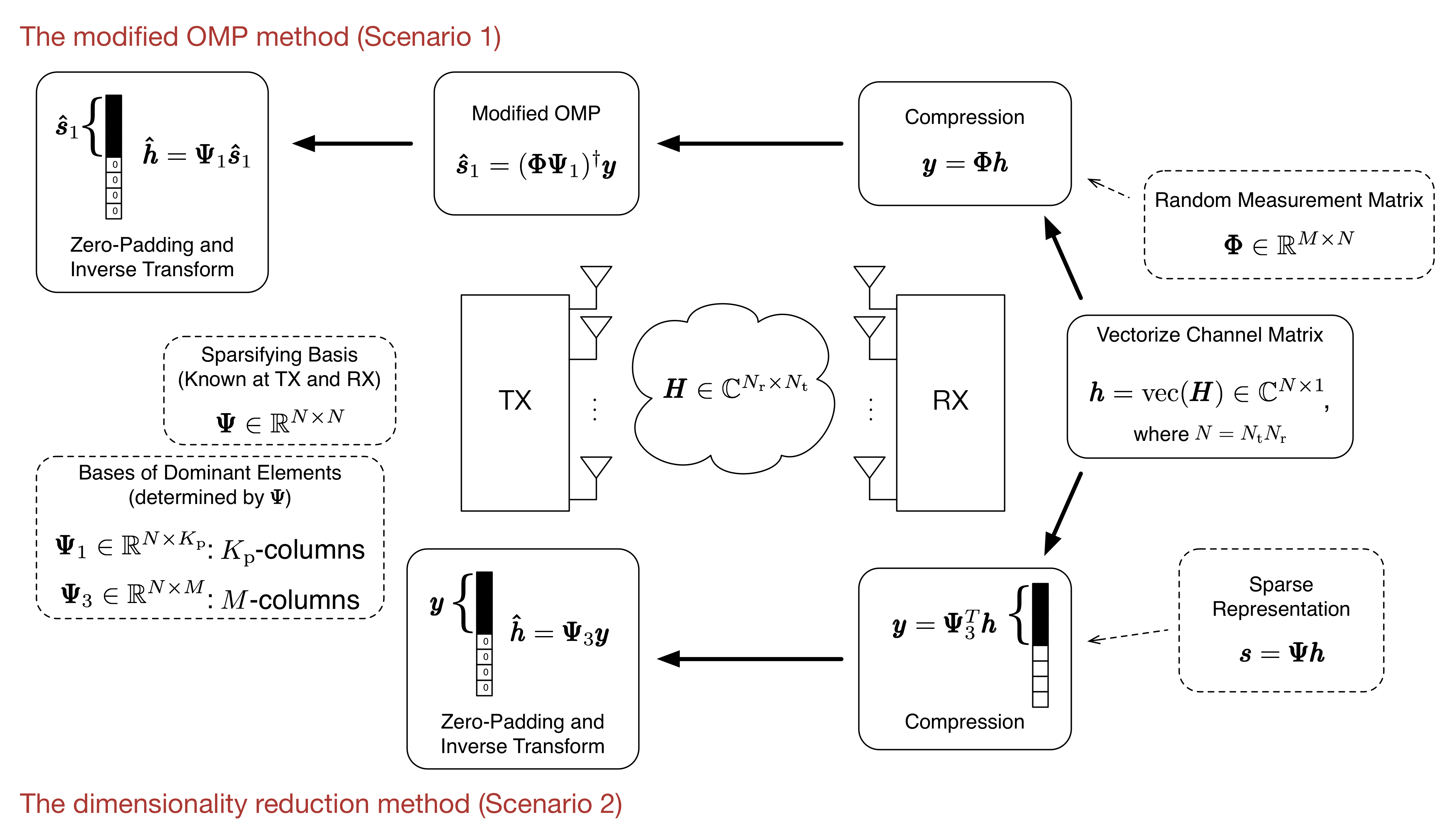}}}
   \caption{A schematic of the proposed MIMO channel feedback methods.}
   \label{Fig:figure_schematic}
\end{figure*}

The two-dimensional UPA ($N_{\mathrm{V}} \times {\mathrm{H}}$) model can be extended from the ULA model~\cite{Adhikary2013JSDM}. For the UPA model, we employ the Kronecker model to express the $N_\mathrm{r} \times N_\mathrm{t}$ spatially-correlated MIMO channel matrix  between the transmitter and the $k$-th receiver:
\begin{equation}
\pmb{H}_{k} = \frac{1}{\sqrt{\mathrm{tr}(\pmb{R}_{\mathrm{RX},k})}} \pmb{R}_{\mathrm{RX},k}^{\frac{1}{2}} \pmb{H}_{\mathrm{iid}} \pmb{R}_{\mathrm{TX},k}^{\frac{1}{2}},\nonumber
\end{equation}
where $\pmb{H}_{\mathrm{iid}}$ is an $N_\mathrm{r} \times N_\mathrm{t}$ matrix whose elements follow the independent and identically distributed (i.i.d.) complex zero-mean, unit variance Gaussian random distribution, and $\pmb{R}_{\mathrm{RX},k}$ and $\pmb{R}_{\mathrm{TX},k}$ are the spatial correlation matrices at the $k$-th receiver and the transmitter. 
To simplify ULA modeling, the correlation matrix $\pmb{R}_{\mathrm{TX},k}$ for the $k$-th receiver is expressed as~\cite{Loyka2001channel}:
\begin{equation}
[\pmb{R}_{\mathrm{TX},k}]_{p,q} = \frac{1}{2\Delta} \int^{\Delta+\phi_k}_{-\Delta+\phi_k} e^{-j2\pi \frac{d}{\lambda} (p - q) sin(\alpha)} d\alpha, \nonumber
\end{equation}
where $\lambda$ is the carrier wavelength, $\Delta$ is the angular spread, and $\phi_k$ is the angle of arrival (AoA) for the $k$-th receiver. The correlation matrix of the UPA model can be expressed by combining the vertical correlation matrix $\pmb{R}_{\mathrm{V}} \in \mathbb{C}^{N_{\mathrm{V}} \times {\mathrm{V}}}$, and the horizontal correlation matrix $\pmb{R}_{\mathrm{H},k} \in \mathbb{C}^{N_{\mathrm{H}} \times {\mathrm{H}}}$ using the Kronecker product $\pmb{R}_{\mathrm{TX},k} = \pmb{R}_{\mathrm{V}} \otimes \pmb{R}_{\mathrm{H},k}$. The angular spread and the AoA for vertical and horizontal correlation matrices are given as~\cite{Adhikary2013JSDM}:
\begin{eqnarray}
\Delta_{\mathrm{V}} &=& \frac{1}{2}\left(\arctan\left(\frac{s+r}{u}\right) - \arctan\left(\frac{s-r}{u}\right)\right), \nonumber \\
\phi_{\mathrm{V}} &=& \frac{1}{2}\left(\arctan\left(\frac{s+r}{u}\right) + \arctan\left(\frac{s-r}{u}\right)\right), \nonumber \\
\Delta_{\mathrm{H}} &=& \arctan\left(\frac{r}{s}\right), \nonumber \\
\phi_{\mathrm{H},k} &\in& (-\pi,\pi], \nonumber
\end{eqnarray}
where $u$, $r$, and $s$ are the elevation of the transmit antenna, the radius of the scattering ring for the receiver, and the distance from the transmitter, respectively. 
For simulations, we set $u=60\mathrm{m}$, $r=30\mathrm{m}$, and $s=100\mathrm{m}$. The channel matrix including all $N_\mathrm{u}$ receivers is formed by stacking, column-wise, the channel matrices between the transmitter and each receiver
\begin{equation}
\pmb{H} = \left[\begin{array}{c}\pmb{H}_1^T~\pmb{H}_2^T~\cdots~\pmb{H}_{N_\mathrm{u}}^T\end{array}\right]^T.\nonumber
\end{equation}

\section {Background}

\label{background}
In this section, we briefly review how CS works and discuss the importance of the original signal's sparsity in reconstruction. To encode sparse signals, several compression methods are available. Some signals have sparsity themselves while others can be sparsified in some domain that makes only a few dominant coefficients sufficient to represent the signals. Consider an $N \times 1$ target signal $\pmb{x}$, which can be sparsified into an $N \times 1$ sparsified signal $\pmb{s}$ with an $N \times N$ sparsifying basis $\pmb{\Psi}$ as
\begin{equation}
\pmb{s} = \pmb{\Psi}^T \pmb{x},\nonumber
\end{equation}
where $\pmb{s}$ has at most only $K$ non-zero elements. This type of signal $\pmb{s}$ is called $K$-sparse. If the target signal $\pmb{x}$ has sparsity itself, the sparsifying basis $\pmb{\Psi}$ can be an identity matrix. The commonly used examples of $\pmb{\Psi}$ include the discrete Fourier transform (DFT) matrix and the discrete cosine transform (DCT) matrix. Since such transformations are usually orthonormal, the target signal can be represented as $\pmb{x} = \pmb{\Psi} \pmb{s}$. In practice, it is hard to expect the sparsified signal $\pmb{s}$ to be sparse. In such a case, $\pmb{s}$ is assumed to be noisy-sparse, which has $K$ dominant elements and $(N-K)$ negligible elements. To compress the target signal $\pmb{x}$, we introduce two methods: 1) CS, and 2) the dimensionality reduction.

\subsection{Compressive Sensing}

The greatest advantage of CS is not needing to know the indices (positions) of the non-zero elements in $\pmb{s}$. With CS, the target signal $\pmb{x}$ is blindly encoded as an $M \times 1$ measurement vector $\pmb{y}$ via random projections as:
\begin{equation}
\label{measure} 
\pmb{y} = \pmb{\Phi} \pmb{x} = \pmb{\Phi} \pmb{\Psi} \pmb{s},
\end{equation}
where $\pmb{\Phi}$ is an $M \times N$ measurement matrix, which can be generated randomly according to the distributions such as Gaussian or Bernoulli. The compression capability is bounded as $M \ge cK \text{log} \frac{N}{K}$ for some small constant $c$ \cite{Candes:rip}, \cite{donoho2006compressed}. The compression ratio $\eta$ is calculated as $\eta=M/N$.

Since $\pmb{\Phi}$ is a wide matrix, $\pmb{y} = \pmb{\Phi} \pmb{x}$ is an undetermined linear system of equations. To reconstruct $\pmb{x}$ from $\pmb{y}$, the decoder solves the following $\ell_1$-norm minimization problem: 
\begin{equation}
\text{min}\|\pmb{s}\|_{\ell_1} \quad s.t. \quad \pmb{y} = \pmb{\Phi} \pmb{\Psi} \pmb{s},\nonumber
\end{equation}
which is typically solved by optimization algorithms such as basis pursuit (BP). The decoder can also reconstruct $\pmb{s}$ by greedy algorithms such as orthogonal matching pursuit (OMP) \cite{tropp2007signal}. The exact reconstruction of $\pmb{x}$ is guaranteed with high probability by the Restricted Isometry Property (RIP) of $\pmb{\Phi} \pmb{\Psi}$ \cite{Candes:rip}.

\subsection{Encoding by Dimensionality Reduction}

It is an intuitive step to compress $\pmb{x}$ by encoding the dominant elements in $\pmb{s}$ by dimensionality reduction. In this case, the information on indices of such elements has to be known at the encoder, and also has to be fed back to the decoder. With some sparsifying basis $\pmb{\Psi}$, however, the position of dominant elements in $\pmb{s}$ is expected to be in certain region. Therefore, the encoder and the decoder can fix the order of encoding/decoding $\pmb{s}$. For example, in image processing, JPEG uses the 2D-DCT as a sparsifying basis and encodes low frequency data priorly.

\section{Massive MIMO Channel Feedback}
\label{feedback}
In this section, we introduce the two-dimensional discrete cosine transform (2D-DCT) and the Karhunen-Loeve transform (KLT) as a sparsifying basis. We also explain how the receiver encodes and feeds back CSI to the transmitter. To reduce feedback overhead, we propose to compress CSI into an $M \times 1$ vector via random projection or dimensionality reduction. With each sparsifying basis, we specify the position of the dominant elements in the sparsified CSI vectors.

\subsection{Sparsifying Basis}
\label{basis}

An efficient sparsifying basis is needed to reconstruct the compressed sparse signal with lower error. In practical cases, the sparsified signal $\pmb{s}$ may have $K$ dominant elements and other $(N-K)$ elements may not be zero, which means it would not be $K$-sparse. Since the reconstruction algorithms assume that the sparsified signal is $K$-sparse, they reconstruct only $K$ elements. Other elements are considered as errors. Therefore, it is important to use an efficient sparsifying basis that makes non-dominant elements smaller. 

To handle CSI easily, $\pmb{H}_{k}$ is vectorized into an $N_\mathrm{r} N_\mathrm{t} \times 1$ vector
\begin{equation}
\pmb{h}_{k} = \text{vec}(\pmb{H}_{k}).\nonumber
\end{equation}
For convenience, we omit the supscript $k$. We design a sparsifying basis $\pmb{\Psi}$ to sparsify $\pmb{h}$. The sparsifying performance of $\pmb{\Psi}$ plays a key role in reconstruction in both compression methods with the fixed compression ratio $\eta = M / (N_\mathrm{r} N_\mathrm{t})$.

\smallskip
\subsubsection{The Two-Dimensional Discrete Cosine Transform Basis}
Due to the spatial correlation among the antennas, $\pmb{H}$ is expected to be sparse in spatial-frequency domain. The 2D-DCT is widely used in lossy compression of audio and images because of its strong energy compaction property and simplicity of computing. Also, if the 2D-DCT is chosen to be used as a sparsifying basis, there is no need to calculate the basis, meaning the basis is fixed. Note that the matrix operation of the 2D-DCT can be written as $\pmb{C}_{N_\mathrm{r}}^T \pmb{H} \pmb{C}_{N_\mathrm{t}}$, where $\pmb{C}_N$ is the $N \times N$ DCT matrix. This can be written in a vector form as: 
\begin{equation}
\pmb{s}_\textrm{DCT} = (\pmb{C}_{N_\mathrm{t}} \otimes \pmb{C}_{N_\mathrm{r}})^T \text{vec}(\pmb{H}) = (\pmb{C}_{N_\mathrm{t}} \otimes \pmb{C}_{N_\mathrm{r}})^T \pmb{h}.\nonumber
\end{equation}
Therefore, a sparsifying basis with the 2D-DCT is $\pmb{\Psi}_\textrm{DCT} = (\pmb{C}_{N_\mathrm{t}} \otimes \pmb{C}_{N_\mathrm{r}})$. An advantage of the 2D-DCT as a sparsifying basis is that $\pmb{\Psi}_\textrm{DCT}$ is fixed even though the correlation of the channel changes. In other words, the receiver and the transmitter need not to calculate $\pmb{\Psi}_\textrm{DCT}$ as the correlation changes. Since 2D-DCT ignores information on how the channel is correlated, however, the sparsifying performance is limited.

\begin{figure}[!t]
        \centering
        \begin{subfigure}[b]{0.6841\columnwidth}
                \centering
                \includegraphics[width=1.0\columnwidth]{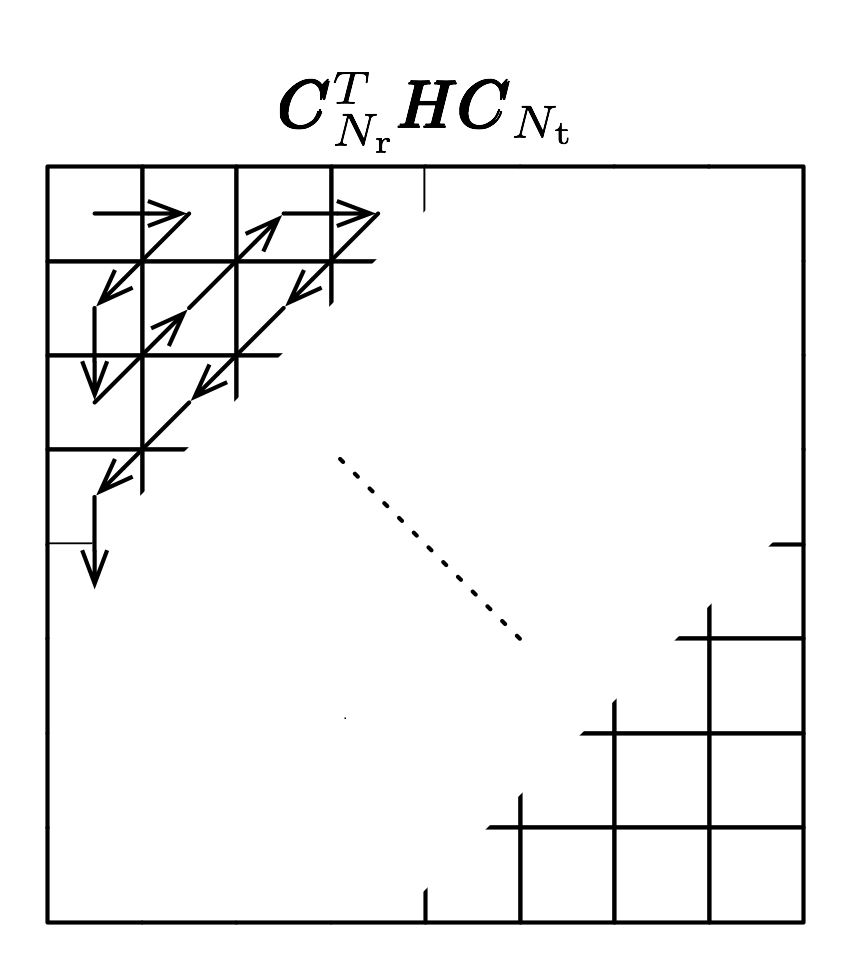}
                \caption{The 2D-DCT basis}
                \label{Fig:comp_order1}
        \end{subfigure}
        \begin{subfigure}[b]{0.2659\columnwidth}
                \centering
                \includegraphics[width=1.0\columnwidth]{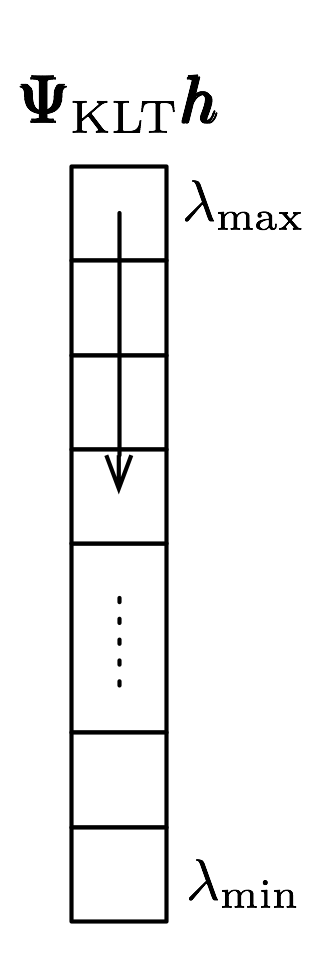}
                \caption{The KLT basis}
                \label{Fig:comp_order2}
        \end{subfigure}
        \caption{The order of selecting dominant elements in sparsified CSI with the sparsifying bases.}        
        \label{Fig:comp_order}
\end{figure}

\begin{figure*}[t]
        \centering
        \begin{subfigure}[b]{0.66\columnwidth}
                \centering
                \includegraphics[width=1.0\columnwidth]{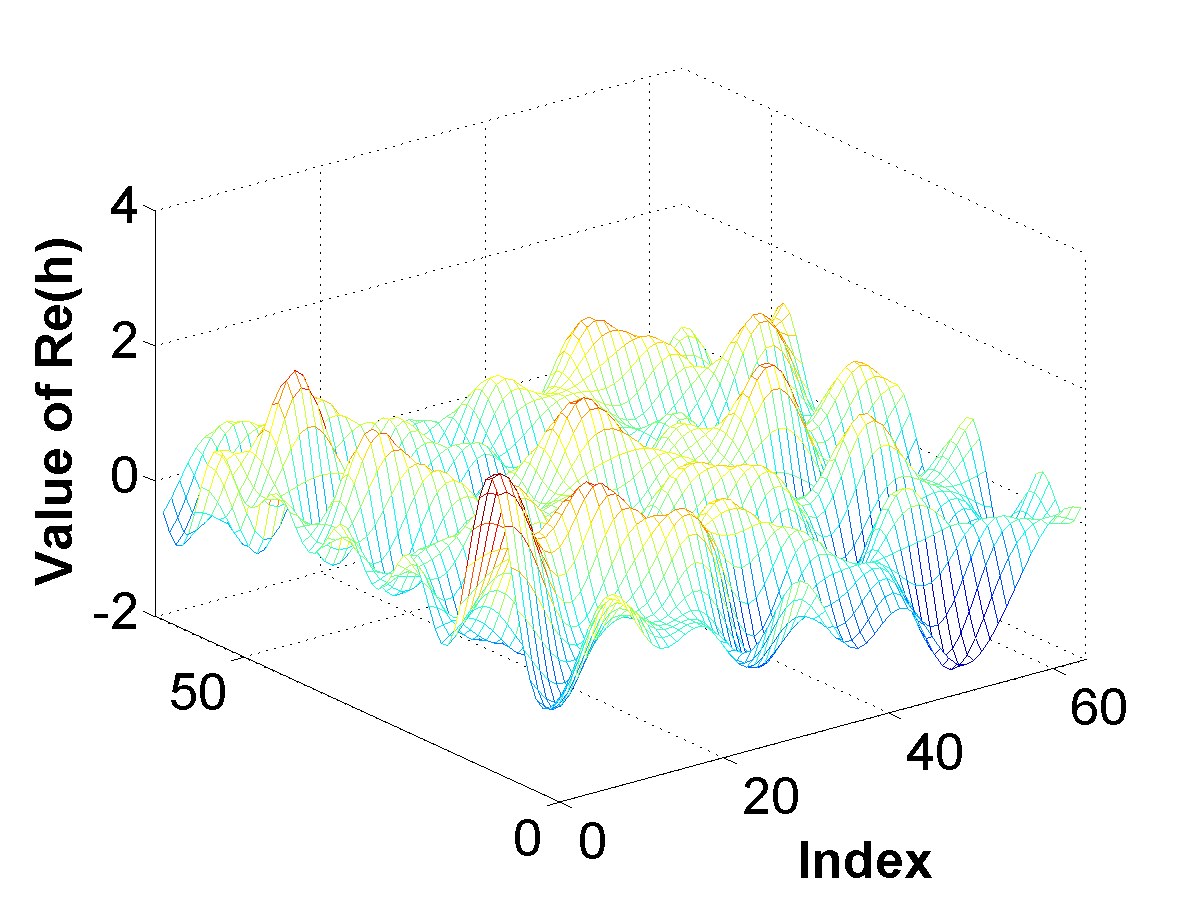}
                \caption{$\text{Re}(\pmb{h})$ with an ULA.}
                \label{linear_H}
        \end{subfigure}
        \begin{subfigure}[b]{0.66\columnwidth}
                \centering
                \includegraphics[width=1.0\columnwidth]{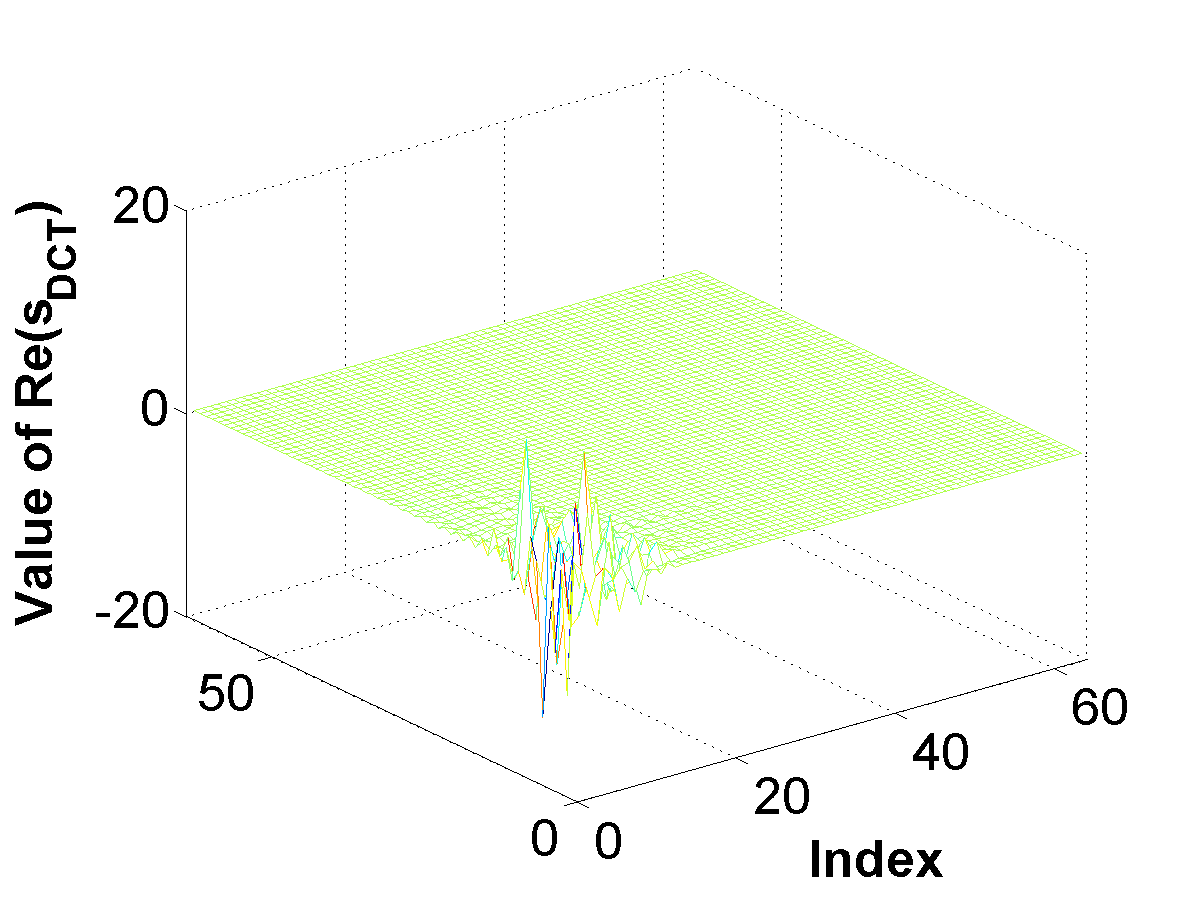}
                \caption[bf]{$\text{Re}(\pmb{s}_\textrm{DCT})$ with an ULA.}
                \label{linear_dct}
        \end{subfigure}
        \begin{subfigure}[b]{0.66\columnwidth}
                \centering
                \includegraphics[width=1.0\columnwidth]{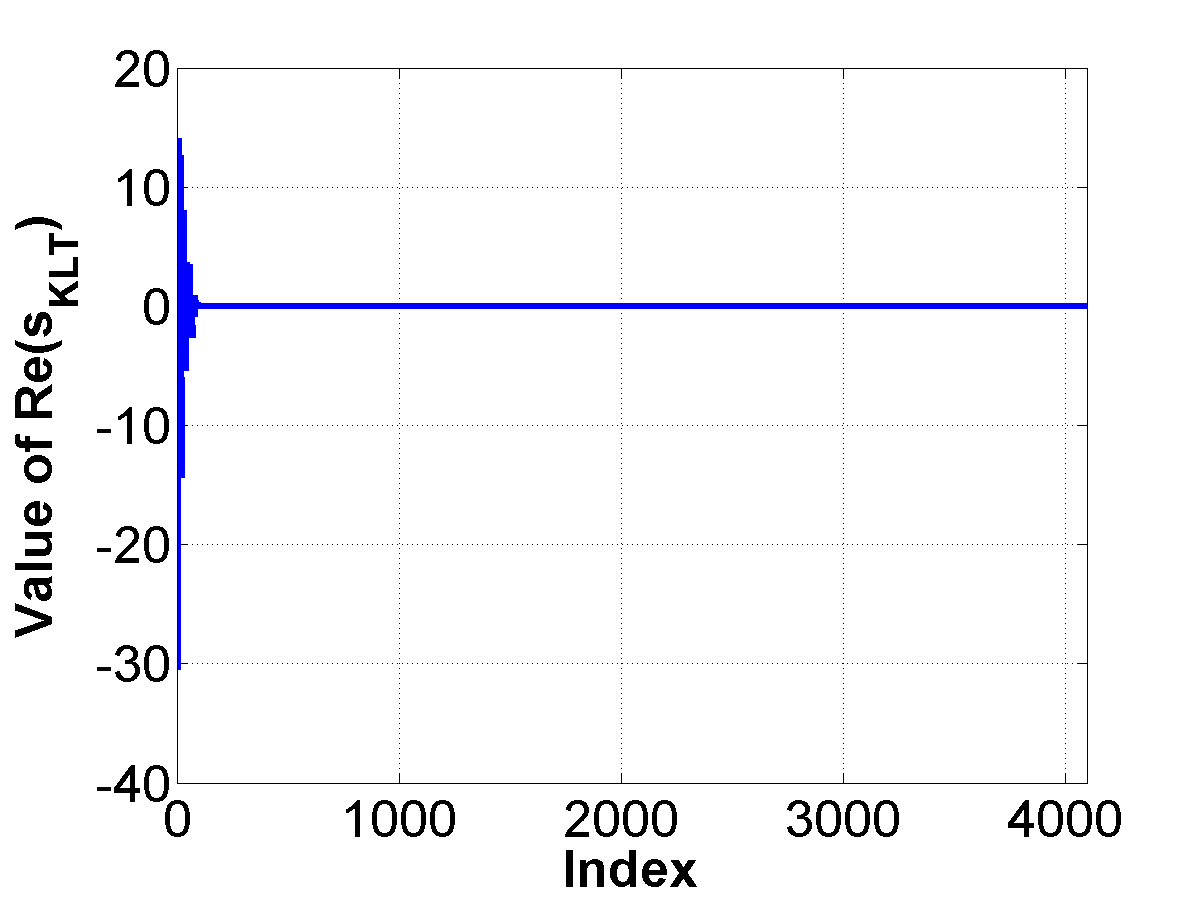}
                \caption{$\text{Re}(\pmb{s}_\textrm{KLT})$ with an ULA.}
                \label{linear_pca}
        \end{subfigure}
        
        \begin{subfigure}[b]{0.66\columnwidth}
                \centering
                \includegraphics[width=1.0\columnwidth]{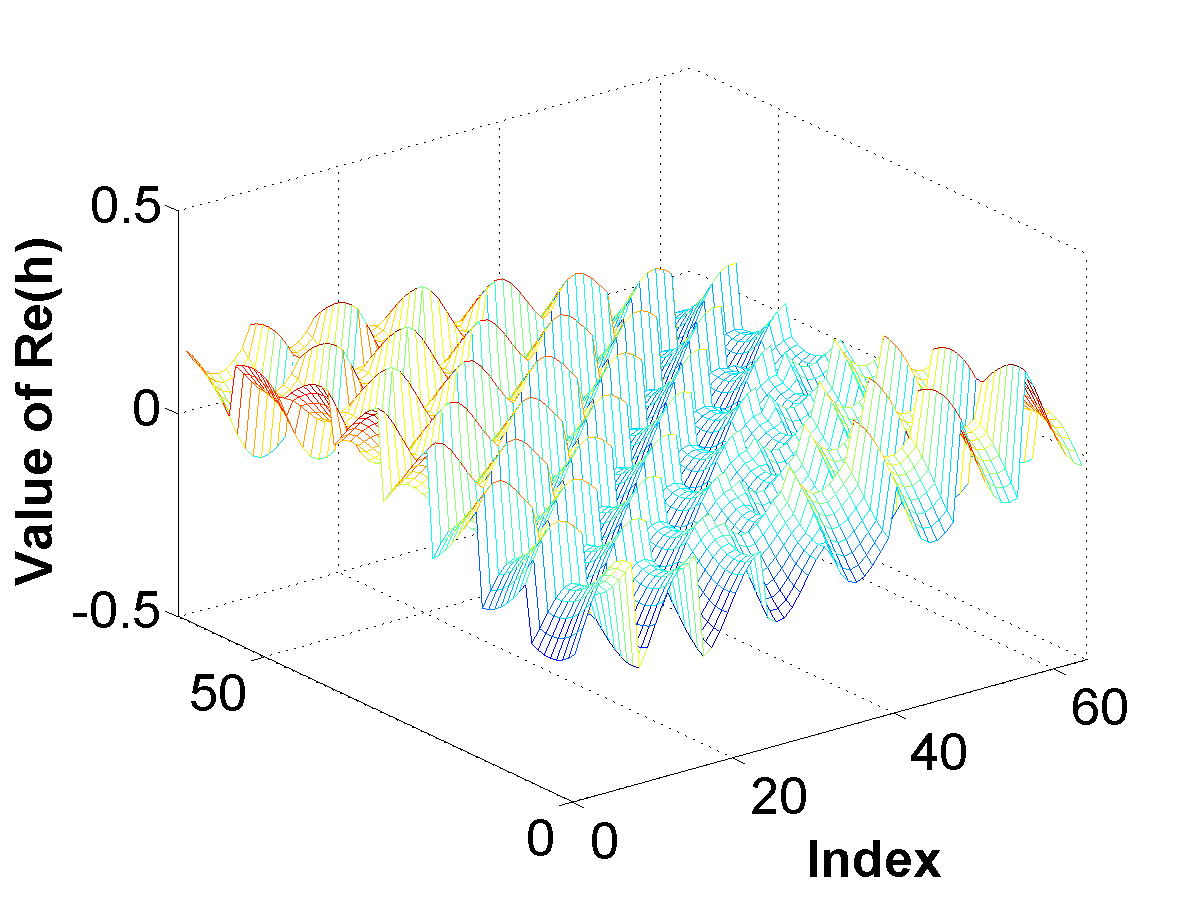}
                \caption{$\text{Re}(\pmb{h})$ with an UPA.}
                \label{square_H}
        \end{subfigure}
        \begin{subfigure}[b]{0.66\columnwidth}
                \centering
                \includegraphics[width=1.0\columnwidth]{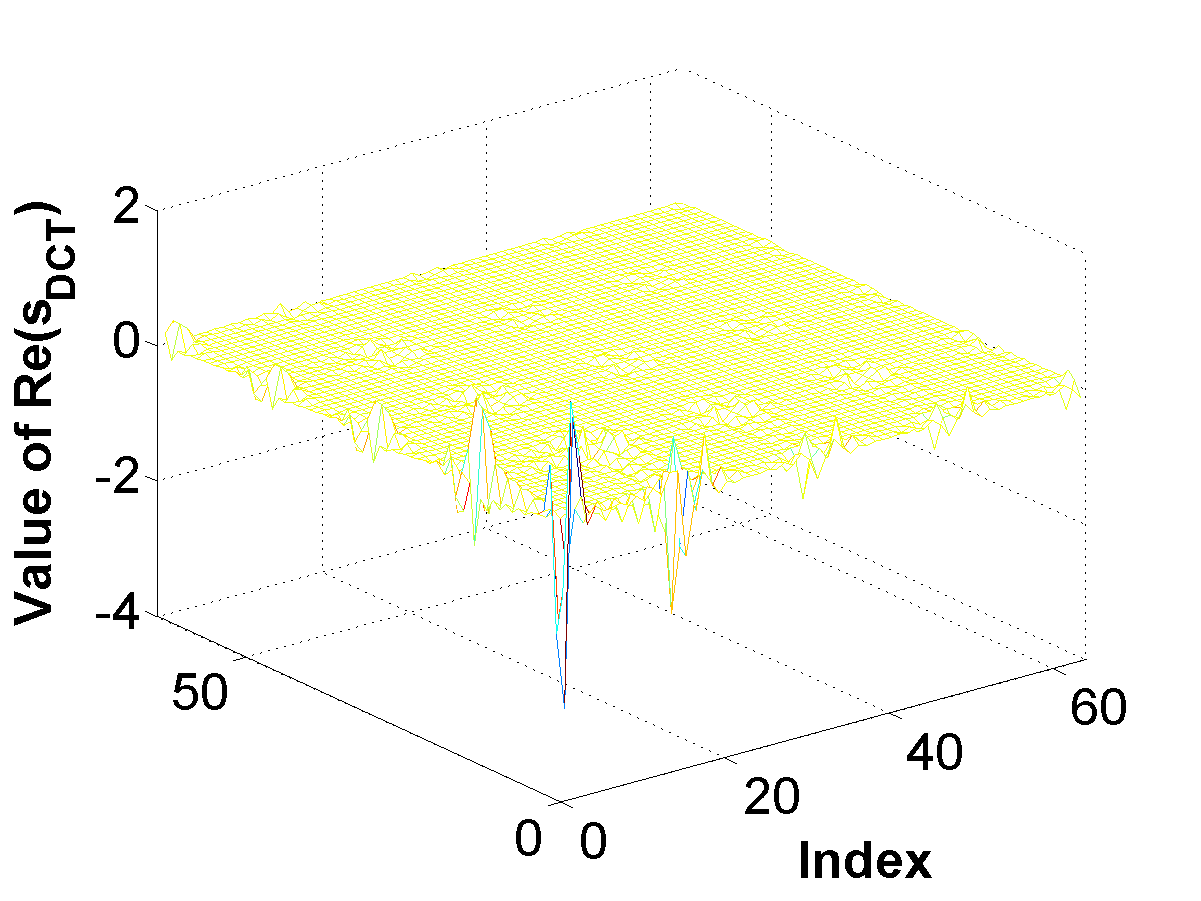}
                \caption{$\text{Re}(\pmb{s}_\textrm{DCT})$ with an UPA.}
                \label{square_dct}
        \end{subfigure}
        \begin{subfigure}[b]{0.66\columnwidth}
                \centering
                \includegraphics[width=1.0\columnwidth]{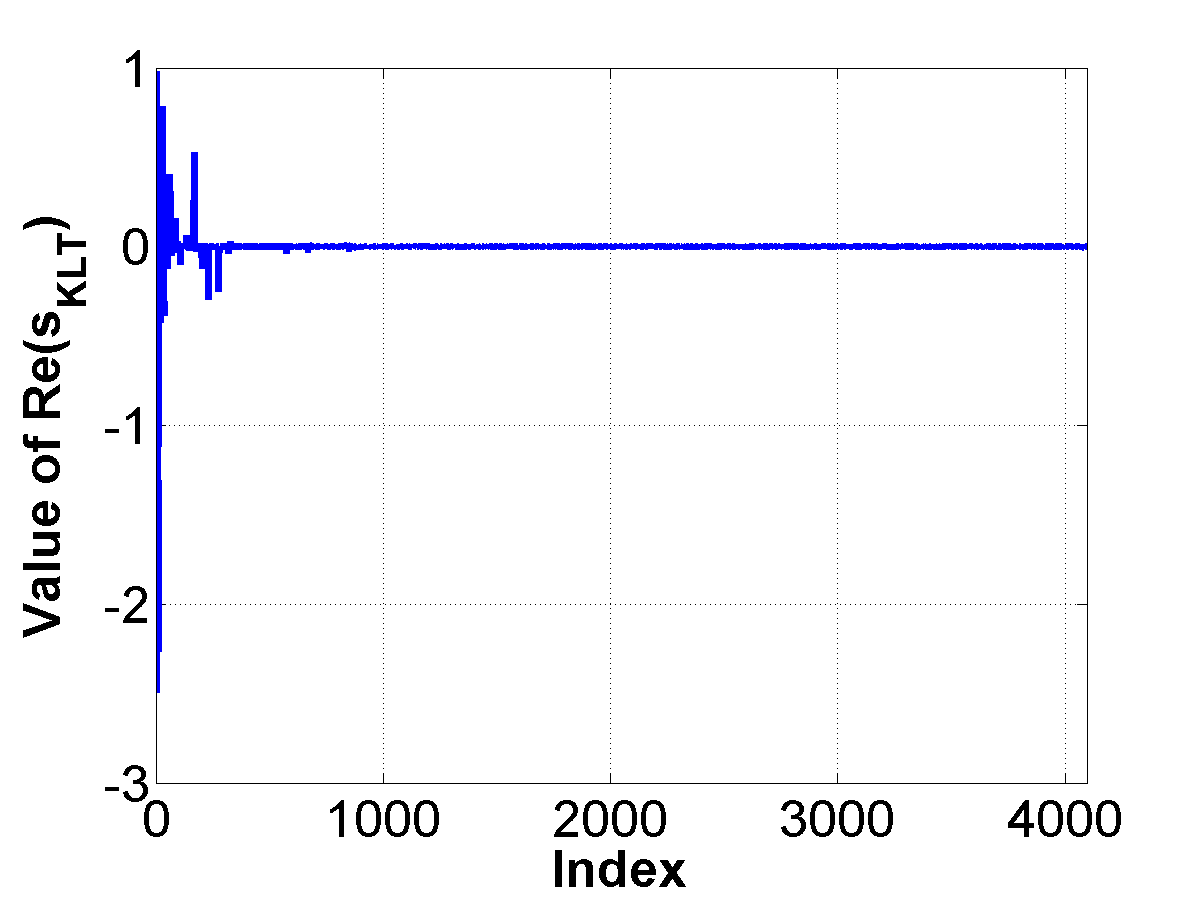}
                \caption{$\text{Re}(\pmb{s}_\textrm{KLT})$ with an UPA.}
                \label{square_pca}
        \end{subfigure}
        \caption{The real part of CSI and sparsified CSI with two bases. (a), (b), (c) are the results with an ULA of antennas, and (d), (e), (f) are the results with an UPA of antennas.}
        \label{Fig:plot_channel}
\end{figure*}

\smallskip
\subsubsection{The Karhunen-Loeve Transform Basis}
We assume that the spatial correlation of the channel is fixed. Therefore, we can employ the sparsifying basis spanned by the eigenvectors of the covariance $\pmb{C}_{\pmb{h}}$ of $\pmb{h}$. The covariance $\pmb{C}_{\pmb{h}}$ can be formulated as
\begin{eqnarray}
\label{covariance} \nonumber
\pmb{C}_{\pmb{h}} &=& \mathbb{E}[\pmb{h} \pmb{h}^*].\nonumber
\end{eqnarray}
Since $\pmb{C}_{\pmb{h}}$ is a Hermitian matrix, the proposed sparsifying basis, $\pmb{\Psi}_\textrm{KLT}$, can be computed by the eigenvalue decomposition: 
\begin{equation}
\label{evd} 
\pmb{C}_{\pmb{h}} = \pmb{\Psi}_\textrm{KLT} \pmb{\Lambda} \pmb{\Psi}^{-1}_\textrm{KLT} = \pmb{\Psi}_\textrm{KLT} \pmb{\Lambda} \pmb{\Psi}^*_\textrm{KLT},\nonumber
\end{equation}
where $\pmb{\Psi}_\textrm{KLT}$ is a matrix consisting of normalized eigenvectors of $\pmb{C}_{\pmb{h}}$, while $\pmb{\Lambda}$ is a diagonal matrix whose elements are corresponding eigenvalues. With the proposed basis $\pmb{\Psi}_\textrm{KLT}$, the covariance of the sparsified CSI vector $\pmb{s}_\textrm{KLT} = \pmb{\Psi}_\textrm{KLT}^* \pmb{h}$ is calculated as
\begin{eqnarray}
\label{cov_s}
\pmb{C}_{\pmb{s}_\textrm{KLT}} &=& \mathbb{E}[\pmb{s}_\textrm{KLT} \pmb{s}_\textrm{KLT}^*] \nonumber = \mathbb{E}[\pmb{\Psi}_\textrm{KLT}^* \pmb{h}  \pmb{h}^* \pmb{\Psi}_\textrm{KLT}]\nonumber \\
&=& \pmb{\Psi}_\textrm{KLT}^* \mathbb{E}[\pmb{h}  \pmb{h}^*] \pmb{\Psi}_\textrm{KLT} = \pmb{\Psi}_\textrm{KLT}^* \pmb{C}_{\pmb{h}} \pmb{\Psi}_\textrm{KLT}  = \pmb{\Lambda}.
\end{eqnarray}
The elements of $\pmb{s}$ are independent of each other, and the variance of the $i$-th element of $\pmb{s}$ is the $i$-th eigenvalue $\lambda_i$. Due to the high correlation in the channel, $\pmb{\Lambda}$ has only a few dominant elements, which means the proposed basis provides the powerful sparsifying performance.

To obtain the KLT basis, the channel covariance $\pmb{C}_{\pmb{h}}$ must be estimated. It is reasonable to assume that $\pmb{C}_{\pmb{h}}$ changes slowly compared to the coherence time of the channel $\pmb{H}$. Furthermore, it is known that $\pmb{C}_{\pmb{h}}$ is frequency invariant for the wide-sense stationary uncorrelated scattering (WSSUS) fading model~\cite{molisch2011wireless}. $\pmb{C}_{\pmb{h}}$, therefore, can be obtained at the transmitter (or BS) by the uplink in FDD systems, or by subspace tracking algorithm~\cite{eriksson1994online} using the downlink training.\footnote{One might argue that it is not true for large frequency separation. In this paper, however, for simplicity, we assume the covariance is equal in different frequency band.}

\subsection{Proposed Channel Compression and Feedback}
\label{channel_comp}

The sparse signal compression techniques explained in Section~\ref{background} can be applied to compress the channel feedback. As mentioned in Section~\ref{intro}, we consider two scenarios. In Scenario~1, since the receiver does not know which basis will be used, the CS technology is adopted, and the CSI is compressed via random projection. In Scenario~2, the receiver knows which sparsifying basis will be adopted. Therefore, CSI can be compressed by dimensionality reduction. 
Figure~\ref{Fig:figure_schematic} shows the schematic of the CS-based and the dimensionality-reduction-based MIMO channel feedback methods.

\smallskip
\subsubsection{Specifying the Indices of Dominant Elements in Sparsified Channel Information}
\label{dominant_elements}

With the sparsifying bases introduced in Section~\ref{basis}, the indices of the dominant elements in sparsified CSI $\pmb{s}$ are expected to be focused on certain region. 2D data from nature such as pictures tend to have the most energy in low frequency. From this observation, the indices of the dominant elements of the sparsified CSI from the 2D-DCT basis can be specified in low frequency. In the case of selecting $K$-dominant elements, $K$ elements are selected in zig-zag order in a matrix form of sparsified CSI, $\pmb{C}_{N_\mathrm{r}}^T \pmb{H} \pmb{C}_{N_\mathrm{r}}$ as illustrated in Figure~\ref{Fig:comp_order}(a). With the KLT basis, the variance of each element of sparsified CSI $\pmb{s}$ is determined by eigenvalues $\pmb{\Lambda}$ of (\ref{cov_s}). By rearranging the columns (eigenvectors) of $\pmb{\Psi}_\textrm{KLT}$, the eigenvaues $\pmb{\Lambda}$ can be ordered in a descending order. To select $K$ dominant elements in sparsified CSI $\pmb{s}$, therefore, the first $K$ elements of $\pmb{s}$ are selected as illustrated in Figure~\ref{Fig:comp_order}(b). 

Figure~\ref{Fig:plot_channel} shows that the proposed selecting order described in Figure~\ref{Fig:comp_order} is reasonable. We generate the channels with $N_\mathrm{t}=64$, $N_\mathrm{r}=64$, $d = 0.1 \lambda$. \mbox{Figure~\ref{Fig:plot_channel}(a)} is the real part of the channel matrix $\pmb{H}$ with an ULA of antennas. \mbox{Figures~\ref{Fig:plot_channel}(b),~\ref{Fig:plot_channel}(c)} are the sparsified forms of $\pmb{H}$ with the 2D-DCT basis, and the KLT basis, respectively. To check that the encoding order is reasonable in Figure~\ref{Fig:comp_order}, the sparsified CSI with the 2D-DCT basis is represented in a matrix form, and the sparsified CSI with the KLT basis is represented in a vector form. \mbox{Figures~\ref{Fig:plot_channel}(d),~\ref{Fig:plot_channel}(e),~\ref{Fig:plot_channel}(f)} show the same things, but with an $8 \times 8$ UPA of antennas on both transmitter and receiver. For convenience, in the rest of this paper, we rearrange the columns of two sparsifying bases, $\pmb{\Psi}_\textrm{DCT}$ and $\pmb{\Psi}_\textrm{KLT}$, so that the first $K$ elements of $\pmb{s}$ are selected as dominant elements.

\begin{table}[t!]
\centering
\caption{OMP algorithm}
\footnotesize\normalsize
\begin{tabular}{  p{2.3cm}  p{5.7cm} }
	\hline \hline
 	Input: & measurements $\pmb{y}$ \\
 	& measurement matrix $\pmb{\Phi}$ \\
 	& sparsifying basis $\pmb{\Psi}$ \\
 	& sparsity $K$ \\
 	Initialize: & iteration coount $k=0$ \\
 	& residual vector $\pmb{r}^0 = \pmb{y}$ \\
 	& estimated support set $T^0 = \phi$ \\
	\hline
	While $k < K$ & \\
	& $k = k+1$ \\
	& $t^k = \displaystyle \arg \max_j |\langle\pmb{r}^{k-1}, \phi_j\rangle|$ \\
	& $T^k = T^{k-1} \cup \{t^k\}$ \\
	& $\pmb{\hat{s}}_{T^k} = \displaystyle \arg \min_{\pmb{x}} \| \pmb{y} - \pmb{\Phi}_{T^k}\pmb{s} \|_2 $ \\
	& $\pmb{r}^k = \pmb{y} - \pmb{\Phi}_{T^k}\pmb{\hat{s}}_{T^k}$ \\
	End & \\
	Reconstruction: & $\pmb{\hat{s}} = \displaystyle \arg \max_{\pmb{x}:\mathrm{supp}(\pmb{x})=T^K} \| \pmb{y} - \pmb{\Phi}\pmb{s} \|_2 $ \\
	Output: & $\pmb{\hat{h}} = \pmb{\Psi} \pmb{\hat{s}}$ \\
	\hline\hline
	Complexity: & $\mathcal{O}(N_\mathrm{r} N_\mathrm{t}MK)$ \\
	\hline\hline
\end{tabular}
\label{Table:OMP}
\end{table}

\smallskip
\subsubsection{Compressive Sensing-based Feedback with Modified OMP (Scenario~1)}

In Scenario~1, the receiver does not know whether the transmitter can exploit the KLT basis. Note that the compression part of CS does not need a sparsifying basis, random projection is used for compressing the CSI. Using random projection for compression, the transmitter can, for reconstruction, use either the 2D-DCT basis or the KLT basis. Since the sparsifying performance of the KLT basis is better than that of the 2D-DCT basis, the transmitter adopts, if it can, the KLT basis.

In CS-based compression, according to (\ref{measure}), the $N_\mathrm{r} N_\mathrm{t} \times 1$ CSI $\pmb{h}$ is encoded into the $M \times 1$ measurement vector $\pmb{y}$ via random projections:
\begin{equation}
\pmb{y} = \pmb{\Phi} \pmb{h} = \pmb{\Phi} \pmb{\Psi} \pmb{s},\nonumber
\end{equation}
where an $M \times N_\mathrm{r} N_\mathrm{t}$ measurement matrix $\pmb{\Phi}$ is generated by the i.i.d. Gaussian distribution with zero-mean, and unit variance, and we assume that both the receiver and the transmitter share $\pmb{\Phi}$. After the transmitter obtains the compressed data $\pmb{y}$, it reconstructs the channel $\pmb{\hat{h}}$. The reconstruction algorithms of CS, including BP and OMP, reconstruct the sparsified signal $\pmb{\hat{s}}$ and multiply $\pmb{\Psi}$ to get $\pmb{\hat{h}} = \pmb{\Psi} \pmb{\hat{s}}$.

OMP is a widely used algorithm due to its low computational comprexity. The complexity can be expressed as $\mathcal{O}(N_\mathrm{r} N_\mathrm{t}MK)$ in a linear funciton of sparsity level $K$ \cite{WeiDai2009subspace}. It iteratively investigates the support of the sparsified signal. In each iteration, the correlation between each column of $\pmb{\Phi} \pmb{\Psi}$ and the modified measurements (so called residual) are compared to identify the elements of the support as explained in Table~\ref{Table:OMP}. OMP, therefore needs $K$-iterations for reconstruction.

The support of dominant elements in $\pmb{s}$, can be specified without iterations, as explained in Section~\ref{dominant_elements}. Since the order of selecting the dominant elements is known, only the number of the dominant elements has to be determined. Let $K_\mathrm{p} (\le M)$ denote the number of the dominant elements to be reconstructed and be determined empirically considering $M$ and sparsity of $\pmb{s}$. In the later section, we suggest some intuition to choose proper $K_\mathrm{p}$. Modified OMP, therefore, is proposed as Table~\ref{Table:MOMP} with complexity of $\mathcal{O}(N_\mathrm{r} N_\mathrm{t}M)$.  The sparsified signal can be represented by the sum of the dominant elements part and the non-dominant part:
\begin{equation}
\pmb{s} = \left[\begin{array}{c}\pmb{s}_1\\ \pmb{0}_1\end{array}\right] + \left[\begin{array}{c}\pmb{0}_2 \\ \pmb{s}_2 \end{array}\right],\nonumber
\end{equation}
where $\pmb{s}_1$ and $\pmb{s}_2$ represent the $K_\mathrm{p} \times 1$ dominant part and the $(N_\mathrm{r} N_\mathrm{t} -K_\mathrm{p}) \times 1$ non-dominant part of the sparsified signal, respectively, and $\pmb{0}_1$ and $\pmb{0}_2$ represent an $(N_\mathrm{r} N_\mathrm{t} -K_\mathrm{p}) \times 1$ zero vector and a $K_\mathrm{p} \times 1$ zero vector, respectively. The sparsifying basis can also be separated as:
\begin{equation}
\pmb{\Psi} = \left[\begin{array}{c}\pmb{\Psi}_1 ~~ \pmb{\Psi}_2\end{array}\right],\nonumber
\end{equation}
where $\pmb{\Psi}_1$ and $\pmb{\Psi}_2$ are an $N_\mathrm{r} N_\mathrm{t} \times K_\mathrm{p}$ and $N_\mathrm{r} N_\mathrm{t} \times (N_\mathrm{r} N_\mathrm{t}-K_\mathrm{p})$ matrix, respectively, which consist of the columns of $\pmb{\Psi}$. Since $K_\mathrm{p} \le M$, the $K_\mathrm{p} \times 1$ reconstructed dominant elements $\pmb{\hat{s}}_1$ is obtained as:
\begin{eqnarray}
\pmb{\hat{s}}_1 &=& (\pmb{\Phi} \pmb{\Psi}_1)^\dagger \pmb{y}\nonumber \\
&=& (\pmb{\Phi} \pmb{\Psi}_1)^\dagger (\pmb{\Phi} \pmb{\Psi}_1 \pmb{s}_1 +\pmb{\Phi} \pmb{\Psi}_2 \pmb{s}_2)\nonumber \\
&=&  \pmb{s}_1 + (\pmb{\Phi} \pmb{\Psi}_1)^\dagger \pmb{\Phi} \pmb{\Psi}_2 \pmb{s}_2, \nonumber
\end{eqnarray}
where $(\cdot)^\dagger$ denotes the Moore-Penrose pseudoinverse. 
The reconstructed CSI $\pmb{\hat{h}}$ is obtained by inverse transform:
\begin{equation}
\pmb{\hat{h}} = \pmb{\Psi}_1 \pmb{\hat{s}}_1. \nonumber
\end{equation}
The squared error of CSI reconstruction is expressed as:
\begin{equation}
\label{mse_momp}
MSE_{\pmb{h}} = \|\pmb{h} - \pmb{\hat{h}}\|^2_2 = \|(\pmb{\Phi} \pmb{\Psi}_1)^\dagger \pmb{\Phi} \pmb{\Psi}_2 \pmb{s}_2\|^2_2 + \|\pmb{s}_2\|^2_2.
\end{equation}

\begin{table}[t!]
\centering
\caption{Modified OMP algorithm}
\footnotesize\normalsize
\begin{tabular}{  p{2.3cm}  p{5.7cm} }
	\hline \hline
 	Input: & measurements $\pmb{y}$ \\
 	& measurement matrix $\pmb{\Phi}$ \\
 	& sparsifying basis $\pmb{\Psi}$ \\
 	& reconstruction parameter $K_\mathrm{p}$ \\
	\hline
	Dominant basis: & $\pmb{\Psi}_1$ consists of $K_\mathrm{p}$ columns of $\pmb{\Psi}$\\
	& selected as in Section~\ref{dominant_elements} \\
	Reconstruction: & $\pmb{\hat{s}}_1 = (\pmb{\Phi} \pmb{\Psi}_1 )^\dagger \pmb{y}$ \\
	Output: & $\pmb{\hat{h}} = \pmb{\Psi}_1 \pmb{\hat{s}}_1$ \\
	\hline\hline
	Complexity: & $\mathcal{O}(N_\mathrm{r} N_\mathrm{t}M)$ \\
	\hline\hline
\end{tabular}
\label{Table:MOMP}
\end{table}

\begin{figure*}[t]
        \centering
        \begin{subfigure}{1.0\columnwidth}
                \centering
                \includegraphics[width=1.0\columnwidth]{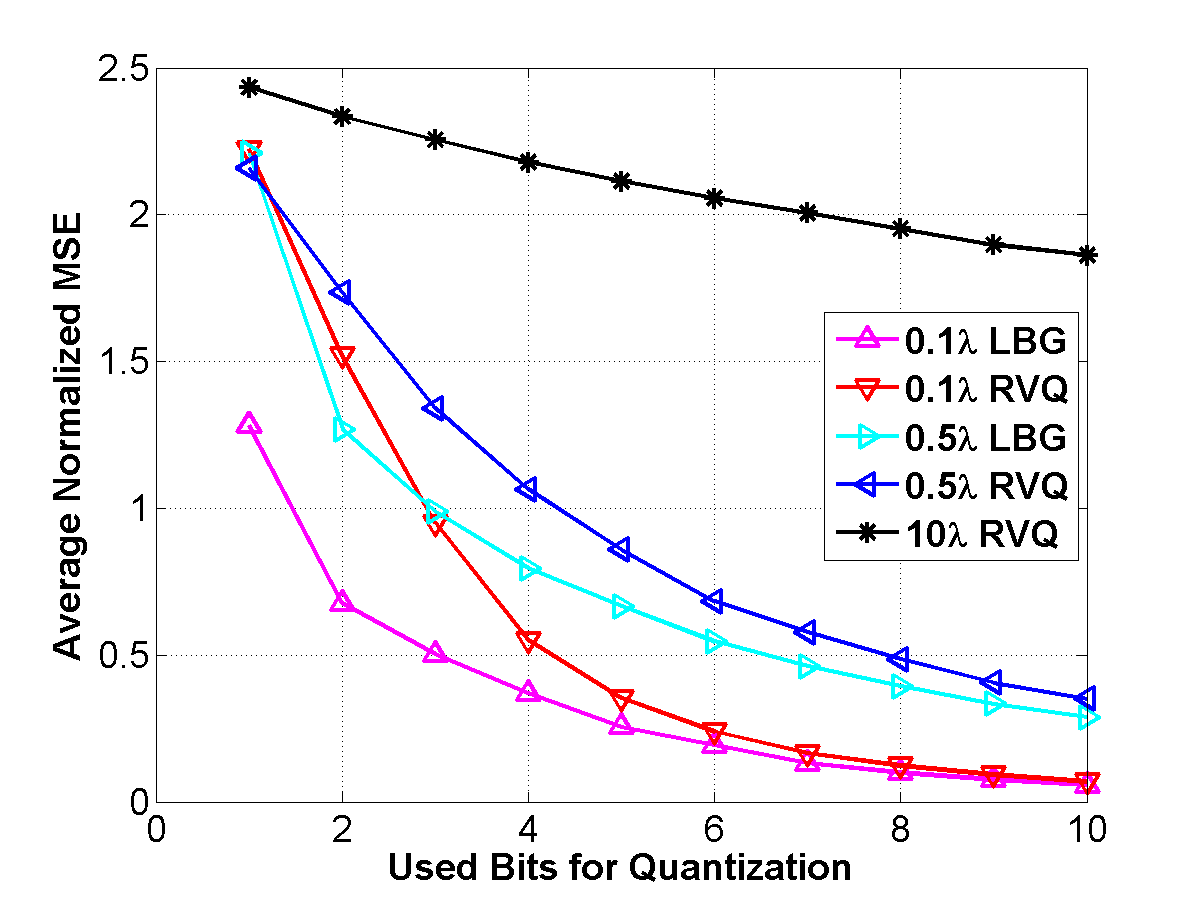}
                \caption{}
                \label{Fig:figure_commse}
        \end{subfigure}        
        \begin{subfigure}{1.0\columnwidth}
                \centering
                \includegraphics[width=1.0\columnwidth]{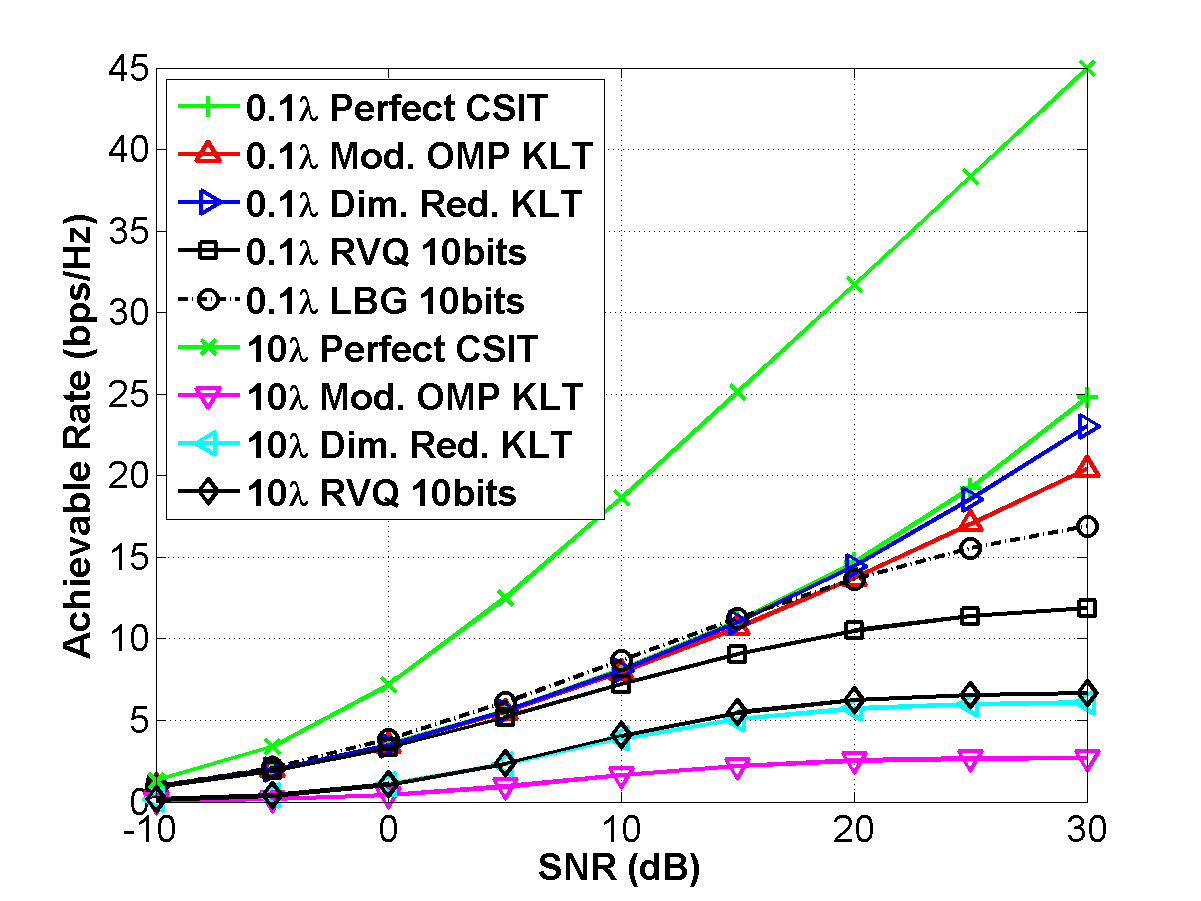}
                \caption{}
                \label{Fig:figure_comrate}
        \end{subfigure}
        \caption{(a) The average normalized MSE of quantized channel feedback with variation of bits used for quantization. (b) Sum rate comparison with perfect CSI, and compressed CSI feedback with $N_\mathrm{u} = 4$. Both simulations are with an $8 \times 8$ UPA of antennas at the transmitter, $N_\mathrm{r}=1$, and $\eta \approx 0.047$.}
        \label{Fig:figure_com}
\end{figure*}

\smallskip
\subsubsection{Dimensionality Reduction based Feedback (Scenario~2)}

In Scenario~2, the receiver knows whether the transmitter can exploit the KLT basis. If both the transmitter and the receiver can exploit the KLT basis, the KLT basis is adopted as a sparsifying basis. If either the transmitter or the receiver cannot exploit the KLT basis, the 2D-DCT basis is used as the sparsifying basis. Since the sparsifying basis $\pmb{\Psi}$ is orthonormal, the $M \times 1$ compressed CSI $\pmb{y}$ and the $N_\mathrm{r} N_\mathrm{t} \times 1$ reconstructed CSI $\pmb{\hat{h}}$ can be obtained:
\begin{eqnarray}
\pmb{y}&=&\pmb{s}_3 = \pmb{\Psi}_3^* \pmb{h}, \nonumber \\
\pmb{\hat{h}}&=&\pmb{\Psi}_3 \pmb{y}, \nonumber
\end{eqnarray}
respectively, where $\pmb{s}_3$ is an $M \times 1$ vector consisting of the first $M$ elements of $\pmb{s}$ and $\pmb{\Psi}_3$ is an $N_\mathrm{r} N_\mathrm{t} \times M$ matrix consists of the first $M$ columns of $\pmb{\Psi}$.

\subsection{Codebook for Compressed Channels}
\label{codebook}
Vector quantization (VQ) \cite{gray1984vector, Gersho1991vector} is a widely used and an efficient technique for data compression. It can be applied for limited feedback in wireless communications. 
The objective of VQ is to represent a set of input vectors $\pmb{v} \in V \subset \mathbb{C}^n$ by a set, $C=\{\pmb{c}_1, \cdots, \pmb{c}_{N_C}\} \subset \mathbb{C}^n$, of $N_C$ code vectors. $C$ is called codebook. VQ can be represented as a mapping:
\begin{equation}
Q:V \rightarrow C. \nonumber
\end{equation}
With the function $Q$, it is possible to define a partition $S$ of set $V$. It is constituted by the encoding region $S_i \subset \mathbb{C}^n$ corresponding to the code vector $\pmb{c}_i$ as:
\begin{equation}
S_i = \{ \pmb{v} \in V | Q(\pmb{v}) = \pmb{c}_i \}. \nonumber
\end{equation}
To evaluate how a vector $\pmb{v}$ is approximated by $\pmb{c}_i$, a distance metric $D$ is defined:
\begin{equation}
D(\pmb{v}, \pmb{c}_i)\equiv \sqrt{(\pmb{v}-\pmb{c}_i)^* (\pmb{v}-\pmb{c}_i)}. \nonumber
\end{equation}
The mean quantization error (MQE) is defined with the fixed codebook $C$ and partition $S$:
\begin{equation}
MQE(C,S) \equiv \mathbb{E}[D(\pmb{v},q(\pmb{v}))]. \nonumber
\end{equation}

For codebook generation, this paper adopts RVQ \cite{AuYeung2007perfomanceRVQ, Santipach2003Asymptotic} and the LBG algorithm \cite{linde1980algorithm}. In RVQ, the codebook $C$, which is known to both the transmitter and receiver, is randomly generated each time the channel changes. The LBG algorithm is an iterative algorithm that uses a training set to solve two optimality criteria which minimize the MQE: the nearest neighbor condition, and the centroid condition.
Given a fixed codebook $C$, the nearest neighbor condition assigns the nearest code vector to each input vector. In other words, the encoding region $S_i$ is obtained by the Voronoi partition~\cite{Gersho1991vector}:
\begin{equation}
S_i = \{\pmb{v} \in V | D(\pmb{v}, \pmb{c}_i) \le D(\pmb{v}, \pmb{c}_j), j \ne i\}. \nonumber
\end{equation}
Given a fixed partition $S$, the centroid condition finds the optimal codebook constituted by the centroid of each encoding region. Therefore, a code vector $\pmb{c}_i$ can be obtained as:
\begin{equation}
\pmb{c}_i = \mathbb{E}[\pmb{v}_i], \quad \pmb{v}_i \in S_i. \nonumber
\end{equation}
A whole iteration of the LBG algorithm obtains \mbox{$(m+1)$-th} codebook $C_{m+1}$ from \mbox{$m$-th} codebook $C_m$ by executing two operations: the calculation of the Voronoi partition of $V$ by adopting the codebook $C_m$; and the calculation of the codebook $C_{m+1}$ whose elements satisfy the centroid condition. This iterative algorithm is repeated until the MQE converges to such value.

\begin{figure*}[t]
        \centering
        \begin{subfigure}{1.0\columnwidth}
                \centering
                \includegraphics[width=1.0\columnwidth]{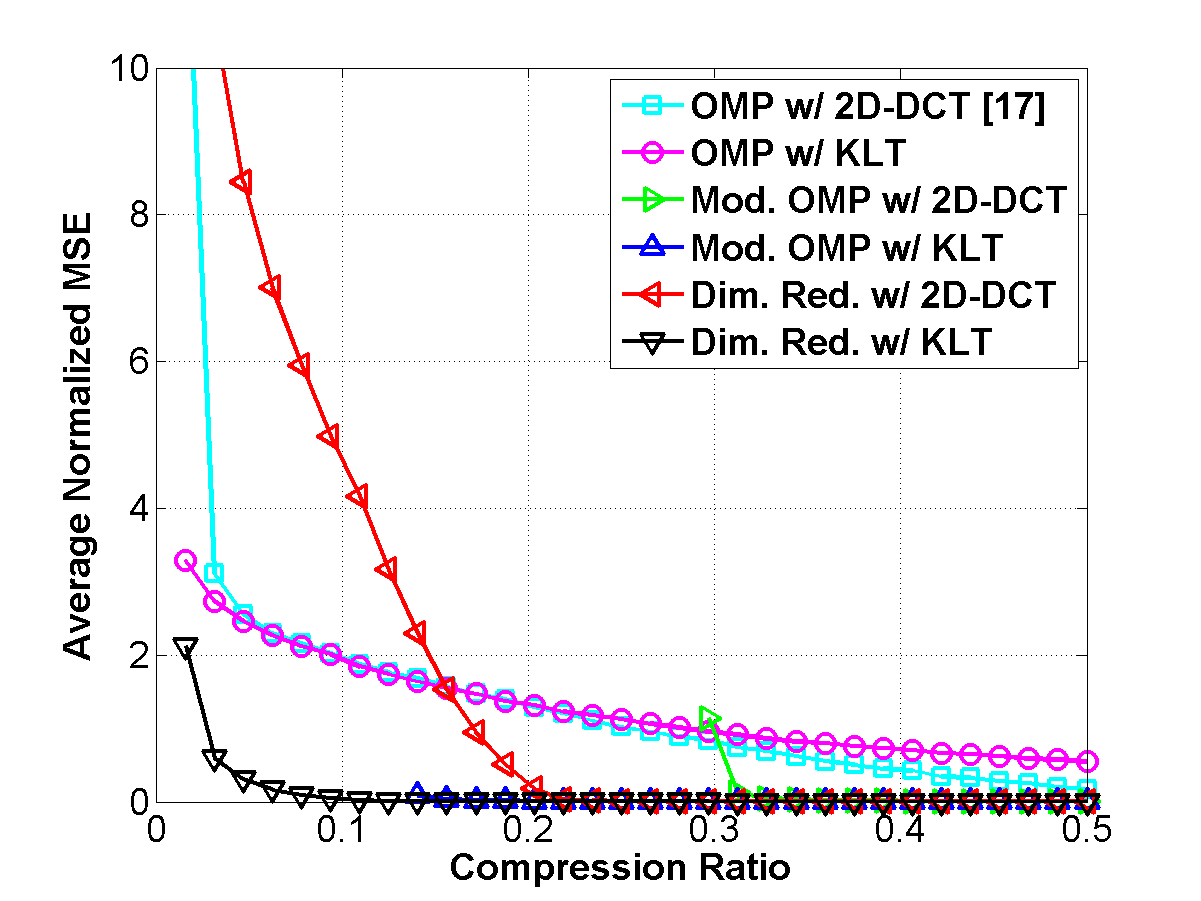}
                \caption{Using an ULA}
                \label{Fig:figure_mse_linear}
        \end{subfigure}               
        \begin{subfigure}{1.0\columnwidth}
                \centering
                \includegraphics[width=1.0\columnwidth]{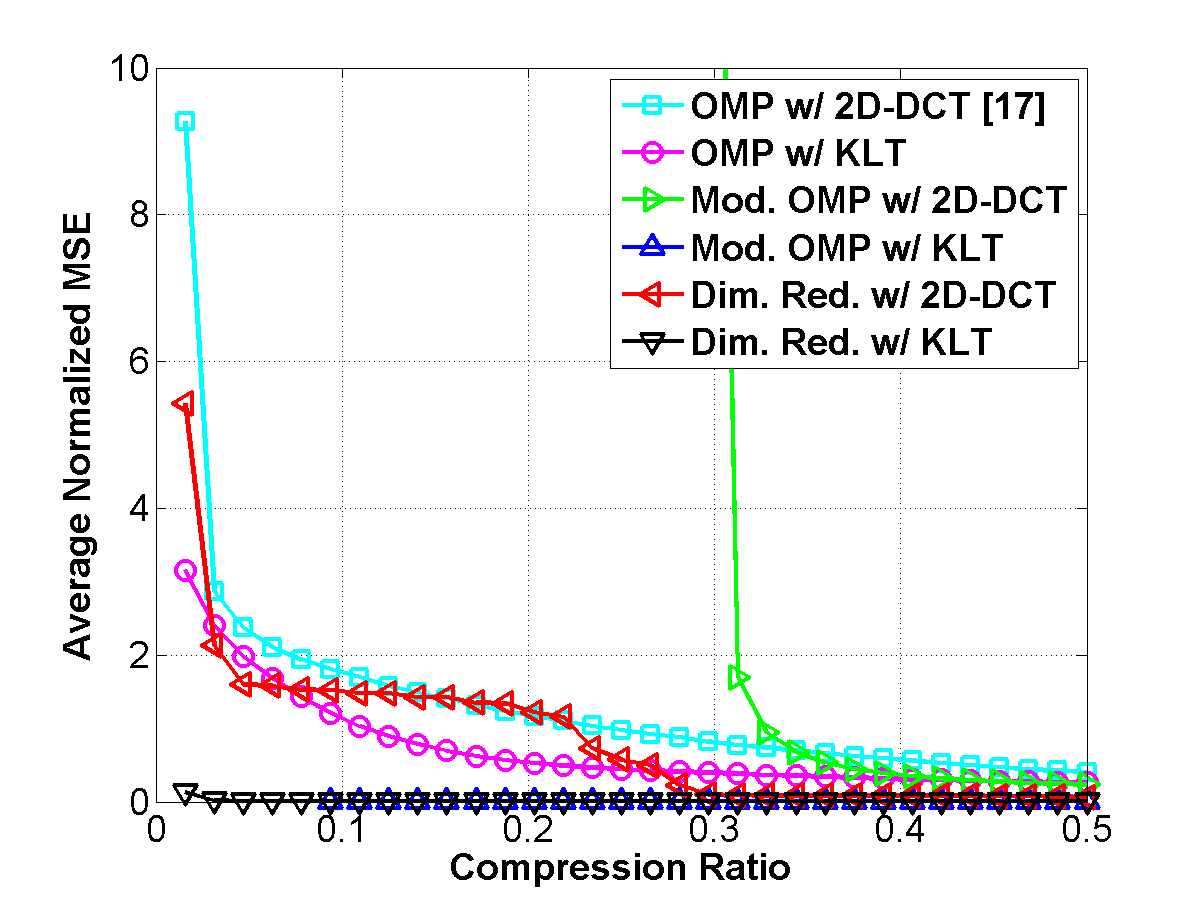}
                \caption{Using an UPA}
                \label{Fig:figure_mse_square}
        \end{subfigure}
        \caption{The average normalized MSE of channel feedback for an ULA and an $8\times8$ UPA with $N_\mathrm{t}=64$, $N_\mathrm{r}=1$, and $d=0.1 \lambda$}
        \label{Fig:figure_mse}
\end{figure*}

After the codebook and the partition are obtained through the iterative part, the splitting part increases, using the obtained codebook, the size of a codebook. The commonly used splitting algorithm doubles the size of codebook by splitting $\pmb{c}_n$ into $(1+\epsilon)\pmb{c}_n$, and $(1-\epsilon)\pmb{c}_n$, where $\epsilon$ is a small constant.
After the splitting part, the iterative part optimizes the codebook and the partition. These two parts of the LBG algorithm are repeated until the desired size of a codebook is obtained.

In this paper, since a CSI vector is to be quantized, the codebook $C$ consists of randomly generated channel vectors $\pmb{h}$ with fixed correlation matrices. To obtain the $b$-bit LBG codebook for the $M \times 1$ compressed feedback vector $\pmb{y}$, two training sequences-sets of input vectors-are generated for two scenarios. For Scenario~1, the training set consists of randomly projected CSI; the training set for Scenario~2 consists of the first $M$ elements of sparsified CSI. Since the LBG algorithm has an optimizing part, it is quite straightforward that the MQE of LBG-based quantization is lower than that of RVQ. The only defect of the LBG algorithm is a need for computing resourse. For the channel whose correlation matrix can be assumed to be static, the quantization performance can be improved through the LBG algorithm.


\section{Performance Analysis}
\label{analysis}
In this section, we justify that, in a massive-MIMO system, the use of highly correlated channels (using small antenna-spacing $d$) outperforms the use of uncorrelated channels (using large $d$). We also compare the performance of three compression methods: the conventional CS-based compression method and the proposed compression methods with different sparsifying bases. 

\subsection{Highly Correlated Channel ($d=0.1\lambda$) vs. Uncorrelated Channel ($d\ge10\lambda$)}
\label{correlated_channel}

It is well-known that higher data rate is achieved with an uncorrelated channel, not a correlated one. If the channel is uncorrelated, however, it is hard to compress, compelling enormous amounts of data to be fed back. Contrarily, a correlated channel can be compressed efficiently, which means the transmitter can obtain more accurate precoding vectors. In summary, a highly correlated channel provides lower achievable rate, but enables the transmitter to exploit better precoding vectors. The channel $\pmb{H}$ only needs enough number of not-close-to-zero singular values (effective rank) to support $N_\mathrm{u}$ receivers. When $N_\mathrm{u}$ is relatively small compared to $N_\mathrm{t}$ and $N_\mathrm{r}$, a correlated channel has enough effective rank to support $N_\mathrm{u}$ receivers. With better precoding performance, a correlated channel can perform a higher sum rate compared to an uncorrelated channel in limited feedback scenarios. 

Figure~\ref{Fig:figure_com} shows the reasonableness of this discussion. We design three types of the transmitters: an $8 \times 8$ UPA of antennas with $d=0.1\lambda$, $d=0.5\lambda$, and $d=10\lambda$. Assume there are 4 receivers and each receiver has one antenna. Each receiver compresses a $64 \times 1$ CSI vector $\pmb{h}$ into a $3 \times 1$ encoded vector $\pmb{y}$ (compression ratio $\eta \approx 0.047$) by the random projection-based and dimensionality-reduction-based compression with the KLT basis. Also, $\pmb{y}$ obtained via dimensionality reduction is quantized with the LBG algorithm, and $\pmb{h}$ is quantized with RVQ. In this paper, we adopt a MMSE precoder. \mbox{Figure \ref{Fig:figure_com}(a)} plots the average normalized mean square error (MSE) of quantized CSI with variation of bits used. It shows that the MSE of quantized CSI with small $d$ is much lower than that with large $d$. \mbox{Figure \ref{Fig:figure_com}(b)} shows that there is only small loss of achievable sum rate when CSI undergoes compression with the correlation of the channel is high, but the loss is big with the less correlated channel. Therefore, a higher sum rate is acheivable with a highly correlated channel with limited feedback.

\subsection{Performance of Single-User MIMO Systems}

The simplest way to compare the performance of the channel feedback is to compare the average normalized MSE between the original $\pmb{h}$ and the fed back $\pmb{\hat{h}}$. We design single-user MIMO systems with a $64$ ULA and an $8 \times 8$ UPA at the transmitter and a single antenna at the receiver with $d=0.1 \lambda$. The reconstruction parameters for the modified OMP are $K_\mathrm{p} = 9$, and $6$ for an ULA and an UPA, respectively, with the KLT basis, and $K_\mathrm{p} = 19$ with the 2D-DCT basis.

Figure~\ref{Fig:figure_mse} shows that CSI is fed back with less error with the KLT basis than with the 2D-DCT basis. Compared to compression error from conventional OMP, both proposed compression methods efficiently decreases compression-error, which means CSI can be compressed with a lower compression ratio. Recalling that both proposed compression methods calls for less computing resourses, we can conclude the proposed methods perform better.
In the UPA case, Figure~\ref{Fig:figure_mse}(b) shows that, when the KLT basis is used, compression-error is lower than the ULA case. Due to the high correlation, the KLT-sparsity of the sparsified CSI is lower with an UPA. 

When 2D-DCT is used for sparsifying, however, there are other dominant elements out of the zig-zag-selected elements. In Figure~\ref{Fig:plot_channel}(e), we can see some extra peaks. These peaks are generated because of an order of antenna indexing in an UPA. As we can see in Figure~\ref{Fig:antenna_array}(b), the correlation or the distance in an UPA does not, unlike an ULA, continually decreases while an index of antenna increases, but increases periodically. For instance, in case of Figure~\ref{Fig:antenna_array}(b), $\rho_{17}$ is larger than $\rho_{13}$. Due to the geometry of an UPA, therefore, the 2D-DCT is not proper for sparsifying UPA CSI.

In Figure~\ref{Fig:figure_mse}, we can observe that MSE of modified OMP using the 2D-DCT basis, compared to others, is abnormally high when $\eta$ is around $0.3$, which means $M \approx K_\mathrm{p}=19$. From the intuition that the condition number of the $M \times K_\mathrm{p}$ matrix $\pmb{\Phi} \pmb{\Psi}_1$ is big when ${M}/{K_\mathrm{p}}$ is close to $1$~\cite{Edelman1988Eigen, Ratnarajah2005Eigen}, we conclude the increase of MSE is due to the pseudoinverse term in~(\ref{mse_momp}). In other words, if $\pmb{\Phi} \pmb{\Psi}_1$ is a square matrix or the numbers of columns and rows are similar, the linear system $\pmb{y}=\pmb{\Phi} \pmb{\Psi}_1 \pmb{s}_1$ becomes sensitive to error term $\pmb{\Phi} \pmb{\Psi}_2 \pmb{s}_2$. $K_\mathrm{p}$, therefore, should not be chosen similar to $M$. For the case using the KLT basis, since the residual error term is small enough, there is no MSE peak.

\begin{figure}[t]
  \centerline{\resizebox{1.0\columnwidth}{!}{\includegraphics{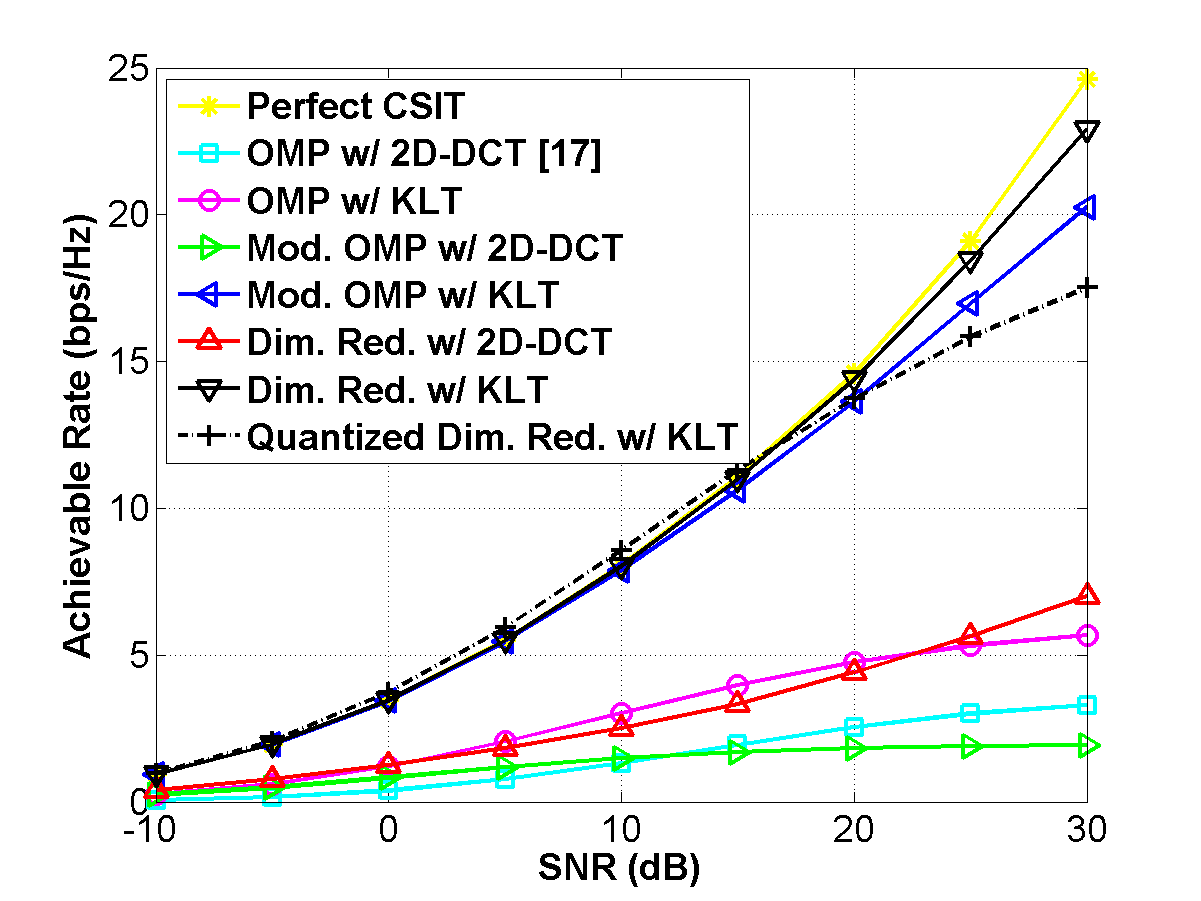}}}
   \caption{Sum rate comparison using a MMSE precoder computed with perfect CSI, compressed CSI by conventional OMP, by the proposed methods, and by the proposed $10$-bit codebook. 
   The system is designed with $N_\mathrm{t}=64$, $N_\mathrm{r}=1$, $N_\mathrm{u}=4$, $d=0.1 \lambda$, and $\eta = 0.047$. An UPA array is implemented on the transmitter.}
   \label{Fig:figure_rate}
\end{figure}

\subsection{Performance of Multi-User MIMO Systems}

The accuracy of the channel feedback in a multi-user system can be measured by the achievable rate, which is the performance of the precoding. In the system of $N_\mathrm{t}=64$, $N_\mathrm{r}=1$, $N_\mathrm{u}=4$, and $d=0.1\lambda$, with an $8\times8$ UPA array, we calculate the sum rate using a MMSE precoder with different channel feedback methods with $\eta = 0.047$. The sum rate with perfect CSI is calculated as the theoretical upper bound, and the sum rate with fed back CSI using conventional OMP with 2D-DCT is calculated as reference data \cite{kuo2012compressive}. 
We compare the sum rates with compressed CSI by three different compression methods: the conventional CS-based methods using either OMP or modified OMP as reconstruction algorithms, and the dimensionality reduction method. Each method is simulated with two kinds of sparsifying bases. Figure~\ref{Fig:figure_rate} shows that the performance of the proposed methods is better than the conventional one. The reconstruction parameters are $K_\mathrm{p} = 3$ and $4$ for modified OMP when the sparsifying bases are the KLT basis and the 2D-DCT basis, respectively. We also simulate with the limited feedback with the 10 bit LBG codebooks.

\section{Conclusions}
\label{conclusion}

This paper proposed sparsifying-based compression mechanisms to reduce the load of the channel feedback in spatially correlated massive MIMO systems. We adopted the KLT basis as sparsifying basis. Using the fact that the indices of the dominant elements in the sparsified CSI, with the particular sparsifying basis, can be specified, we proposed modified OMP for a reconstruction algorithm of CS, and dimensionality-reduction-based compression. For the limited feedback, we applied the LBG algorithm to generate a codebook. We suggested that using highly correlated channels could maximize achievable data rates better than using uncorrelated channels considering the accuracy of the channel feedback in massive MIMO systems. Future work will consider practical issues such as finding a proper reconstruction parameter $K_\mathrm{p}$, correlation estimation, and quantization errors.

\renewcommand{\baselinestretch}{1.0}
\bibliographystyle{IEEEtran}
\renewcommand{\baselinestretch}{1.0}
\bibliography{references_TVT}

\end{document}